# Full characterization of an all fiber source of heralded single photons

Yunxiao Zhang[1], Liang Cui[1,†], Xueshi Guo[1,‡], Wen Zhao[1], Xiaoying Li[1], Z. Y. Ou[2]

1. The State Key Laboratory of Precision Measurement Technology and Instruments, College of Precision Instrument and Opto-Electronics Engineering, Tianjin University, Tianjin 300072, P. R. China
2. Department of Physics, City University of Hong Kong, 83 Tat Chee Avenue, Kowloon, Hong Kong, P. R. China

† lcui@tju.edu.cn;  ‡ xueshiguo@tju.edu.cn

**Abstract**

We demonstrate a heralded single photon source which is based on the photon pairs generated from pulse pumped spontaneous four wave mixing in a piece of commercially available dispersion shifted fiber. The single photon source at 1550 nm telecom band is characterized with both photon counting technique and homodyne detection method. The heralding efficiency and mode purity can be measured by photon counting while the vacuum contribution part can be found by homodyne detection.

## Introduction

A single-photon state is an energy excitation quanta of an optical mode. The sources of single photons are a fundamental resource for quantum optics and quantum information processing (QIP) [1-3]. Generally speaking, a single photon source can be evaluated from the three aspects: i) emission efficiency, described by the probability of emitting one single photon in response to an external request; ii) photon statistical properties, characterized by the second-order correlation function and iii) indistinguishability, related to the mode purity of the single photons. According to the working principal, single photon sources can be classified into two categories: the deterministic sources and the probabilistic sources. The former ones are based on single emitter systems, such as single atoms and quantum dots etc.. The latter ones, also known as the heralded single photon source (HSPS), take advantage of the signal and idler photon pairs generated from spontaneous parametric emission processes, including the parametric down conversion in $\chi^{(2)}$-based crystals and four wave mixing (FWM) in $\chi^{(3)}$-based optical waveguide. For the HSPSs, the signal (or idler) photons are heralded as single photons by the detection event of their counterparts. Because of the high efficiency and simple apparatus, HSPSs are widely employed in implementing the QIP tasks.

Here we demonstrate an all fiber source of heralded single photons, which is pumped by a pulse train originated from a compact mode-locked fiber laser [4-7]. The central wavelength of the HSPS is at 1550 nm telecom band. We previously characterized the fiber sources of HSPSs by using photon counting technique [4,5]. In practice, however, because of the mode impurity and losses, what we have is usually a mixture of vacuum state and a pure single-photon state. Photon counting technique is not responsive to vacuum and cannot distinguish photon in different modes. Homodyne detection, on the other hand, is sensitive to vacuum and depends on mode match between the input and the local oscillator (LO) [8,9]. In this paper, in addition to improving the directly measured efficiency of the HSPS, we will demonstrate how we can measure these imperfections of a real single-photon state generated in our lab and fully characterize it.

## Experiments and Results

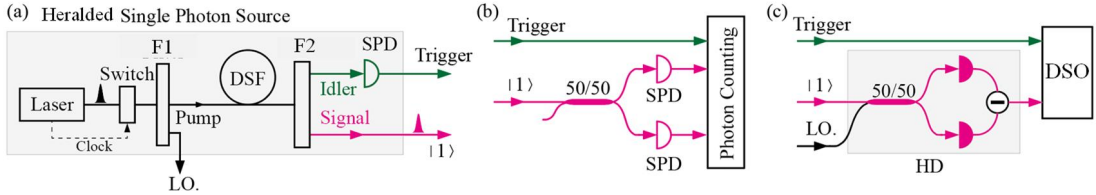

**Fig. 1 Experimental setup for (a) Generation of two-photon state and heralded single-photon state and for characterizing the single photon state |1⟩ with (b) intensity correlation measurement and (c) Homodyne detection (HD), in which a strong local oscillator (LO) serves as a reference.** DSF, dispersion shifted fiber; F1, dual-band filter with pass bands at pump and LO fields; F2, dual-band filter with pass bands at signal and idler fields; SPD, single photon detector; DSO, digital storage oscilloscope.

The experimental setup of the heralded single photon source is shown in Figure 1(a). The heralded single photon state is generated from the $\chi^{(3)}$ nonlinearity in 300-m long dispersion shifted fiber (DSF) through the pulse-pumped spontaneous four wave mixing [4,6,7]. The DSF is cooled to 2.1 K by a cryostat to suppress the background noise contributed by spontaneous Raman scattering [6,7]. The signal and idler photon pairs produced by spontaneous FWM process are normally quite random so photons in each individual field have a thermal statistical distribution. But by gating on the detection of the idler photon, the signal photon is projected into a single-photon state |1⟩. The signal and idler photon pairs are efficiently selected by passing the output of DSF through a dual band filter F2, which is realized by cascading dense wavelength division multiplexing (DWDM) filters [10]. The central wavelengths of pulsed pump, heralded single photon and heralding idler fields are 1549.3, 1553.3 and 1545.3 nm, respectively and the FWHM of the three fields are 0.6, 1.1 and 0.6 nm respectively. The detection events of single photon detector (SPD) placed in idler channel are used as trigger signals to herald the presence of single photons in the signal channel. We measure the heralded signal photon by either direct photon counting (see Fig. 1(b)) or homodyne detection via mixing with a strong local oscillator (LO) (see Fig. 1(c)).

In the experiment, the pulsed pump of HSPS and LO of homodyne detection (HD), centering at 1549.3 and 1553.3 nm, respectively, are created by passing the output of a mode locked fiber laser through a dual-band filter (F1) [*4-6,11*]. The central wavelength, FWHM and repetition rate of the laser are about 1550 nm, 60 nm, and 37 MHz, respectively. So the LO is synchronized with HSPS. Moreover, when the amplitude probability histogram is measured, we need to ensure the response time of our high efficiency HD ($T_R \approx 50$ ns) is fast enough to resolve the single photon events between two adjacent pulses [8]. To do so, the repetition rate of the laser is reduced to 5.3 MHz, which is realized by chopping the laser output with an EOM (electro-optic modulator) based optical switch so that the repetition rate of laser is decreased by 7 times.

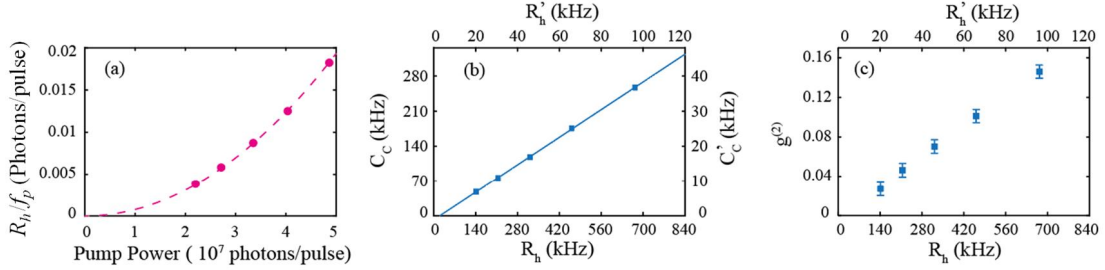

**Fig. 2 Measurement of heralding efficiency and photon statistics of the heralded single photon source.** (a) The detection probability $R_h/f_p$ of heralding idler photons as a function of pump power. (b) Coincidence counting rate $C_c$ ($C_c'$) between the heralded signal channel and heralding idler channel versus heralding rate $R_h$ ($R_h'$) when the pump repetition rate is $f_p \approx 37$ MHz ($f_p' \approx 5.3$ MHz). (c) Intensity correlation $g_h^{(2)}$ of the heralded single photon source versus heralding rate. The dashed curve in plot (a) is the fitting function $R_h = s_1 P_a^2$ with $P_a$ denoting the average of pump photons per pulse. The solid line in plot (b) is the linear fit of the data points.

We first measure the heralding efficiency, photon statistics and mode number of the heralded single photon source by using photon counting technique. Heralding efficiency $\eta_h$ is measured by using two fiber-coupled superconducting nanowire single photon detectors (SPDs) to detect the signal and idler photons when the pump power is varied. The detection efficiency of each SPD is about 80% and its dark counts is less than 500 counts per second. Note that in this measurement, the switch in Fig. 1(a) is either "off" or "on", which corresponds to the pump pulse train with repetition rate of about 37 MHz and 5.3 MHz, respectively. For clarity, we respectively label the heralding rate and coincidence rate of signal and idler photon pairs as $R_h$ ($R_h'$) and $C_c$ ($C_c'$) for the pump with repetition rate of $f_p \approx 37$ MHz ($f_p' \approx 5.3$ MHz). Figure 2(a) shows the detection probability $\frac{R_h}{f_p} (= \frac{R_h'}{f_{p'}})$ of the heralding idler photon as a function of pump power. The data well fits the function $R_h/f_p = s_1 P_a^2$ with $P_a$ denoting the average of pump photons per pulse, which indicates that the heralding photons via FWM are overwhelmingly dominant and the back ground noise from Raman scattering is negligibly small. Figure 2(b) plots the coincidence of two SPDs as a function of the counting rate $R_h$. The data points well fit a solid line, showing the ratio between the coincidence rate $C_c$ and heralding rate $R_h$ is about 0.4. Taking the detection efficiency of SPD (80%) in heralded single photon channel into account, we obtain a heralding efficiency of $\eta_h \approx 50\%$. Figure 2(c) shows the normalized intensity correlation function $g_h^{(2)}$ of the heralded SPS versus the heralding rate $R_h$. Here, $g_h^{(2)}$ is measured by sending the heralded single photon state in the signal field (triggered by the detection of an idler photon) through HBT interferometer consisting of a 50/50 beam splitter and two SPDs (see Fig. 1(b)). The value of $g_h^{(2)}$ increases with heralding rate due to the influence of multi-photon events. But even if the heralding rate is up to 800 KHz, the value $g_h^{(2)} = 0.14 \pm 0.006$ is still well below the classical Poissonian light limit of 1. When the counting rate in heralding idler channel is about $3.4\times10^5$ Hz ($4.85\times10^4$ Hz) for pump repetition rate of $f_p \approx 37$ MHz ($f_p' \approx 5.3$ MHz), we have $g_h^{(2)} = 0.07 \pm 0.006$.

We next measure the mode number of the single-photon source. Considering the thermal nature of individual field of spontaneous parametric processes, we can obtain the mode number of heralded signal field $M_s \approx 1.3$ from the relation $g_s^{(2)} = 1 + \frac{1}{M_s}$ by sending the field into the HBT interferometer and measuring its intensity correlation function $g_s^{(2)} = 1.75$ without the triggering signal from the idler photon detection. Since the mode number of the heralding field sets the lower bound of the modal purity of HSPS, its straight forward to deduce the mode number of HSPS, $M_s \leq 1.3$.

We then calibrate the single photons by using homodyne detector (HD) to measure the quadrature amplitude $\hat{X} = \hat{a}e^{-i\varphi} + \hat{a}^\dagger e^{i\varphi}$, where $\varphi$ is the phase of the strong LO. The size of the electrical current pulses out of HD is proportional to the quadrature amplitude of the measured field, i.e., $\hat{\imath}_{HD}(t) \propto k(t)|\mathcal{E}|\hat{X}(\varphi)$, where $k(t)$ is the time response function of HD and $|\mathcal{E}|$ is the filed amplitude of LO. A single photon state $|1\rangle$ is inherently non-Gaussian and its amplitude probability distribution is symmetric to $\varphi$, so the calibration result is insensitive to the LO phase. Ideally, the measured result obeys probability distribution of $P_1(x) = (2x^2/\pi)\exp(-x^2)$ [8], where $x$ refers to the operator $\hat{X}$ measured by HD. In practice, there inevitably exists the contribution of vacuum induced by non-ideal heralding efficiency and non-ideal HD detection. For overall detection efficiency $\eta$, the amplitude probability distribution can be modeled by

$$P(x) = \eta P_1(x) + (1-\eta)P_0(x), \qquad (1)$$

where $P_0(x) = (2/\pi)\exp(-x^2)$ denotes the amplitude probability distribution for vacuum state $|0\rangle$.

In the process of measuring $P(x)$, the heralded single photon source is directly measured by HD. The HD consists of two identical photo diodes. Before measuring the quadrature amplitude of HSPS, we first evaluate the response function of HD ($k(t)$) by launching the attenuated LO pulses into one photo diode of HD when both the input and the other photo diode in HD are blocked. Figure 3(a) plots the shape of one electrical pulse, showing the raising edge of $k(t)$ is sharp and the FWHM is about 50 ns, but the falling edge will last to about 150 ns. In addition, the oscillation at ~37 MHz originated from the repetition rate of the mode-locked fiber laser is observable in Fig. 3(a) because the optical switch used to reduce the repetition rate of laser by 7 times only provides a limited isolation ( ~10 dB). To effectively evaluate the quadrature amplitude of each pulse, we record the current pulse in time domain and extract its maximum value within the timing window of around 50 ns, as illustrated by the gray shadow area in Fig. 3(a).

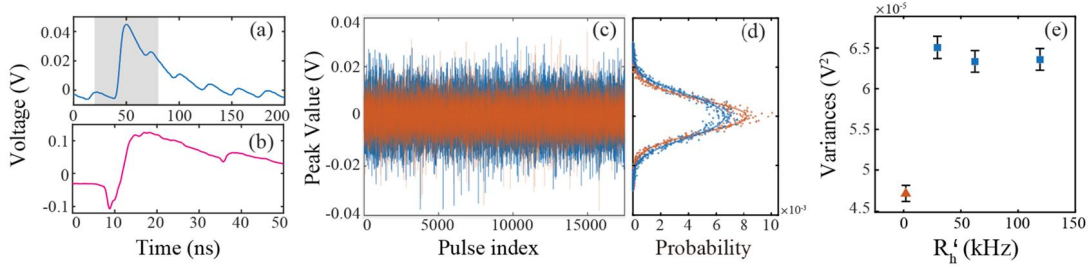

**Fig.3 Results for measuring the amplitude probability distribution of heralded single photo source by homodyne detection (HD).** (a) Normalized response function of one photo diode of homodyne detector. (b) The typical trace of one electronic pulse out of the balanced HD. (c) Peak values of 17500 pulses out of HD for the input field of vacuum (orange trace) and the heralded single photons (blue trace), respectively. (d) The probability histogram of HD output electrical pulses for input of vacuum state (orange dots) and heralded single photons (blue dots). The solid curves are the theoretical fitting of the histogram. (e) Variances of the probability histogram for vacuum input (orange triangle) and heralded single photons (blue squares) when the heralding rate $R_h'$ of single photon source is varied. All the variances of the histogram for SPS with heralding rate $R_h'$ is about 1.2 dB higher than that for vacuum state

We then increase the LO power to about $1.12 \times 10^9$ photons/pulse and measure the quadrature amplitude of input by the HD. To match the mode of HSPS, the central wavelength and FWHM of LO are 1553.3 and 1.1 nm, respectively. The output of HD is converted into voltage and sent to a digital storage oscilloscope (DSO) trigged by the heralding signal of the single photon events. Fig. 3(b) plots a typical trace of one electronical pulse out of HD, in which the height of the peak is proportional to the value of quadrature amplitude of input field. For each set of data, $1.75 \times 10^4$ pulses are recorded and processed. We extract the peak of each pulse and correct it by subtracting the mean value for all the peaks of the recorded electronic current pulses. Fig. 3(c) plots the values of processed peaks when the input of HD is heralded single photon state $|1\rangle$ (blue) and vacuum state $|0\rangle$ (orange), respectively. By performing probability histogram analysis to these peak values (represented by the dots in Fig. 3(d)), we find the variance of the histogram for *H*SPS is about 1.2 dB higher than that for vacuum state. Fitting the histogram for single photon state $|1\rangle$ in Fig.3(d) with Eq. (1), we found the overall detection efficiency is $\eta = 19.2\%$. Considering heralding efficiency of SPS $\eta_h \approx 50\%$, the quantum efficiency of the HD system $\eta_{HD} = 95.0\%$ and non-ideal the transmission efficiency of 90%, mode-matching efficiency between LO and single photon state is deduced to be ~45%. Finally, we repeat the measurement by changing the pump power of HSPS to vary the trigger rate, as shown by the blue squares in Fig. 3(e). We find that all the variances of the histogram for HSPS under different heralding rate $R_h{'}$ is about 1.2dB higher than that for vacuum state (represented by the orange triangle in Fig. 3(e)).

**Conclusion**

In conclusion, we have developed and experimentally demonstrated an all fiber source of heralded single photons at 1550 nm telecom band. We characterize the HSPS by using photon counting technique to measure the heralding efficiency $\eta_h$, the photon statistics $g^{(2)}$ and mode number $M_s$. The results show that we have $\eta_h \approx 50\%$, $M_s \leq 1.3$ and $g^{(2)} \approx 0.07 \pm 0.006$ when the heralding rate is about $3.4 \times 10^5$ Hz. Additionally, we characterize the single photon source by measuring its amplitude probability histogram using homodyne detection with strong LO. We find the variance of the histogram for our HSPS is about 1.2 dB higher than that for vacuum state $|0\rangle$, showing the amplitude of the heralded single photon state $|1\rangle$ can be effectively measured. The performance of the HSPS can be further improved by replacing the nonlinear medium of single piece DSF with a multi-stage nonlinear interferometer [5]. Such an all fiber device of single photon source provides convenience for exploring the nature of light and for studying quantum information technology [12-15].